\documentclass[superscriptaddress,twocolumn,showpacs,prl,floatfix]{revtex4}

\usepackage{color}
\usepackage{tabularx}
\usepackage{epsfig}
\usepackage{amsmath}
\usepackage{natbib}

\usepackage{graphicx}

\renewcommand{\log}{\ln}

\begin{document}

\title{The Ising Spin Glass in dimension four}

\author{P. H.~Lundow} 
\affiliation{Department of Mathematics and Mathematical Statistics,
  Ume{\aa} University, SE-901 87, Sweden}

\author{I. A.~Campbell}

\affiliation{Laboratoire Charles Coulomb (L2C), UMR 5221
  CNRS-Université de Montpellier, Montpellier, F-France
}

\begin{abstract}
The critical behaviors of the bimodal and Gaussian Ising spin glass
(ISG) models in dimension four are studied through extensive numerical
simulations, and from an analysis of high temperature series expansion
(HTSE) data of Klein {\it et al.} (1991). The simulations include
standard finite size scaling measurements, thermodynamic limit regime
measurements, and analyses which provide estimates of critical
exponents without any consideration of the critical temperature. The
higher order HTSE series for the bimodal model provide accurate
estimates of the critical temperature and critical exponents. These
estimates are independent of and fully consistent with the simulation
values.  Comparisons between ISG models in dimension four show that
the critical exponents and the critical constants for dimensionless
observables depend on the form of the interaction distribution of the
model.
\end{abstract}


\maketitle

\section{Introduction}
Renormalization Group Theory (RGT) for thermodynamic phase transitions
\cite{wilson:75} and the Edwards-Anderson model for Ising Spin Glasses
(ISGs) \cite{edwards:75} were introduced almost simultaneously forty
years ago. Ever since it has been tacitly assumed as self-evident that
the standard RGT universality rules should apply to ISGs. As far as we
know there is no rigorous theoretical proof that this ISG hypothesis
holds, though confirmations have been reported a number of times based
on numerical data
\cite{bhatt:88,katzgraber:06,hasenbusch:08,jorg:08}. The universality
principle states that for all systems within a universality class the
critical exponents are strictly identical and do not depend on the
microscopic parameters of the model. All ISG models in a given
dimension are supposed to be in the same universality class, on the
assumption that the form of the interaction distribution is an
irrelevant microscopic parameter. 

Thus in the family of simple ferromagnets, within a universality class
of models having space dimension $d$ and spin dimensionality $n$, all
models have identical critical properties corresponding to an isolated
fixed point in the renormalisation group flow. However, diluted
ferromagnets of given $d, n$ have a different (dilution independent)
set of critical exponents, with values which correspond to a separate
isolated fixed point \cite{ballesteros:98,calabrese:03}.  For a few
special cases of spin models in dimension two (discussed for instance
in Ref.~\cite{cardy:87}) the critical behavior is more complicated
and corresponds to a line of fixed points rather than an isolated
fixed point; critical parameters vary continuously according to motion
along the line, produced by a marginal operator.
From the empirical ISG data below there appear for the moment to be
two possible scenarios : two classes of ISGs (such as models with
continuous distributions and those with discrete distributions) or
alternatively ISG exponents which vary continuously with a parameter
such as the kurtosis of the interaction distribution.  In any case it
has been stated by authoritative authors that \lq\lq classical tools of
RGT analysis are not suitable for spin glasses\rq\rq{}
\cite{parisi:01,castellana:11,angelini:13} although no explicit
theoretical predictions have been made so far concerning the important
question of universality.

Here we combine numerical simulation and high temperature series
expansion (HTSE) data on the bimodal and Gaussian ISGs in dimension
four so as to obtain accurate and reliable values for the critical
parameters in this model. We discuss a number of different methods for
exploiting numerical data, and show that for each model these are
consistent.  Comparisons between these and other estimates on ISG
models in the same dimension but with different interaction
distributions show that the critical exponents and the critical
constants for dimensionless parameters depend on the form of the
interaction distribution.

\section{Numerical techniques}
The Hamiltonian is as usual
\begin{equation}
  \mathcal{H}= - \sum_{ij}J_{ij}S_{i}S_{j}
  \label{ham}
\end{equation}
with the near neighbor symmetric distributions normalized to $\langle
J_{ij}^2\rangle=1$.  The Ising spins live on simple hyper-cubic
lattices with periodic boundary conditions.  We have studied the
bimodal model with a $\pm J$ interaction distribution and the Gaussian
interaction distribution model.  We will compare with published
measurements on these and other $4$d ISGs. We will use the inverse
temperature $\beta = \langle J_{ij}^2\rangle^{1/2}/T = 1/T$, with the
normalization above, or alternatively $w=\tanh^2(\beta)$, to signify
the temperature.  The spin overlap parameter is defined by
\begin{equation}
  q =  \frac{1}{N} \left\langle\sum_{i}S_{i}^{A}S_{i}^{B}\right\rangle
\end{equation}
where $A$ and $B$ indicate two copies of the same system and $N$ is
the number of sites.

The simulations were carried out using the exchange Monte Carlo method
for equilibration using so called multi-spin coding, on $2^{14}$ (up
to $L=7$) or $2^{13}$ (for larger $L$) individual samples at each
size. An exchange was attempted after every sweep with a success rate
of at least 30\%. At least 40 temperatures were used forming a
geometric progression reaching down to $\beta_{\max}=0.55$ for the
bimodal model and $\beta_{\max}=0.60$ for the Gaussian model.
This ensures that our data span the critical
temperature region which is essential for the FSS fits. Near the
critical temperature the $\beta$ step length was at most $0.03$. The
various systems were deemed to have reached equilibrium when the
sample average susceptibility for the lowest temperature showed no
trend between runs. For example, for $L=12$ this means about $200000$
sweep-exchange steps.

After equilibration, at least $200000$ measurements were made for each
sample for all sizes, taking place after every sweep-exchange step.
Data were registered for the energy $E(\beta,L)$, the correlation
length $\xi(\beta,L)$, and for the spin overlap moments.  In addition
the correlations $\langle E(\beta,L),U(\beta,L)\rangle$ between the
energy and observables $U(\beta,L)$ were also registered so that
thermodynamic derivatives could be evaluated using the relation
$\partial U(\beta,L)/\partial \beta = \langle
U(\beta,L)\,E(\beta,L)\rangle-\langle U(\beta,L) \rangle\langle
E(\beta,L)\rangle$ where $E(\beta,L)$ is the energy
\cite{ferrenberg:91}.  Bootstrap analyses of the errors in the
derivatives as well as in the observables $U(\beta,L)$ themselves were
carried out.

\section{Finite size scaling}
\begin{figure}
  \includegraphics[width=3.5in]{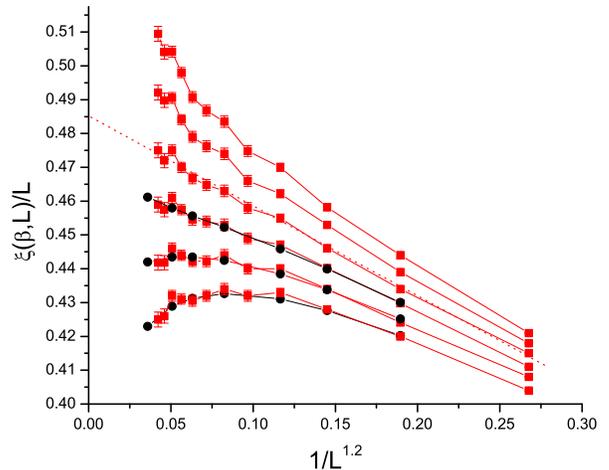}
  \caption{(Color online) The $4$d bimodal ISG. FSS for the
    correlation length ratio $\xi(\beta,L)/L$ at fixed inverse
    temperatures $\beta = 0.5100$, $0.5075$, $0.5050$, $0.5025$,
    $0.5000$,$0.4975$ from top to bottom, against $1/L^{1.2}$. Red
    squares : present data, black circles : read from
    Ref.~\cite{banos:12}. Dashed line : critical
    behavior. }\protect\label{fig:1}
\end{figure}

\begin{figure}
  \includegraphics[width=3.5in]{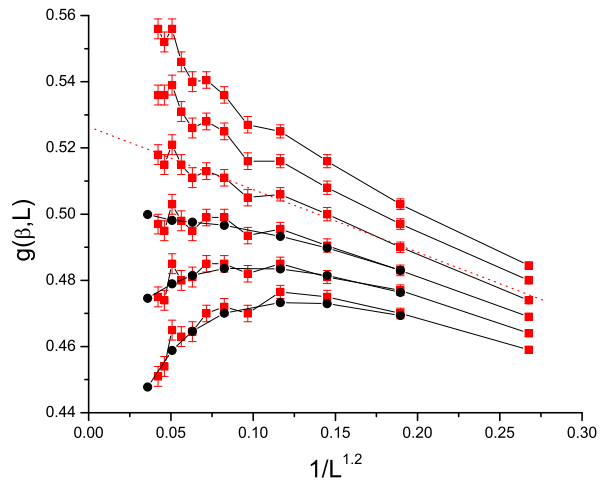}
  \caption{(Color online) The $4$d bimodal ISG. FSS for the Binder
    cumulant $g(\beta,L)$ at fixed inverse temperatures $\beta =
    0.5100$, $0.5075$, $0.5050$, $0.5025$, $0.5000$, $0.4975$ from top
    to bottom, against $1/L^{1.2}$. Red squares : present data, black
    circles : read from Ref.~\cite{banos:12}. Dashed line : critical
    behavior. }\protect\label{fig:2}
\end{figure}

ISG simulations are much more demanding numerically than are those on,
say, pure ferromagnet transitions with no interaction disorder.  The
traditional approach to criticality in ISGs has been to study the
temperature and size dependence of dimensionless observables,
principally the Binder cumulant $g(\beta,L)$ and the correlation
length ratio $\xi(\beta,L)/L$, in the near-transition region and to
estimate the critical temperature and exponents through finite size
scaling (FSS) relations after taking means over large numbers of
samples. Finite size corrections to scaling must be allowed for
explicitly which can be delicate as the range of sizes $L$ is
generally small.  On this FSS approach the estimated values for the
critical exponents are very sensitive to the critical inverse
temperature $\beta_{c}$ estimates. Here we first obtain estimates for
$\beta_{c}$ and the critical parameters using FSS.

The data for standard dimensionless observables, the Binder cumulant
\begin{equation}
  g(\beta,L)=\frac{1}{2}\left(3-\frac{\lbrack\langle q^{4} \rangle\rbrack}{\lbrack\langle q^{2}\rangle\rbrack^2}\right)
  \label{g}
\end{equation}
and the correlation length ratio $\xi(\beta,L)/L$ are shown in
Figs.~\ref{fig:1} and \ref{fig:2} as functions of size at fixed
temperatures near $\beta_{c}$. We show data from the present
simulations together with data taken from the appropriate figures in
Ref.~\cite{banos:12}.  The standard FSS expression for a dimensionless
observable $U=U(\beta,L)$, valid in the critical region is
\begin{equation}
  U = U_c+ AL^{-\omega}+B(\beta-\beta_{c})L^{1/\nu}\left(1+CL^{-\omega} +\cdots\right)
  \label{FSS}
\end{equation}
where $U_c=U(\beta_c,\infty)$. We plot $g(\beta,L)$ and
$\xi(\beta,L)/L$ against $1/L^{\omega}$ with $\omega=1.2$. The choice
of this value, close to the estimate $\omega=1.04(10)$ of
Ref.~\cite{banos:12}, will be explained below.  By inspection there is
good point by point consistency between the present data and those of
Ref.~\cite{banos:12} where the statistical accuracy was much higher,
but where the temperatures studied did not quite span the critical
temperature. The consistency between the two data sets implies that
full equilibrium has been reached. The data in Figs.~\ref{fig:1} and
\ref{fig:2} for fixed $\beta$ should tend to straight lines at
criticality, curving upwards and downwards respectively at large $L$
for $\beta$ higher and lower than $\beta_{c}$; one can estimate
$\beta_{c}=0.505(1)$ from this criterion. The value is marginally
higher than the estimate $\beta_{c}=0.5023(6)$ of
Ref.~\cite{banos:12}.  Data we have obtained on other dimensionless
parameters lead to very similar estimates for $\beta_{c}$
\cite{lundow:14}.

It is important to consider the value of the irrelevant scaling field
thermal correction exponent $\theta$ or equivalently the finite size
correction exponent $\omega = \theta/\nu$.  The first term in the RGT
$\epsilon$-expansion for the ISG leading irrelevant operator is
$\theta(d) = (6-d)$ \cite{dedominicis,bray}, so $\theta(3) \approx 3$,
$\theta(4) \approx 2$, $\theta(5) \approx 1$ (see
\cite{ballesteros:97} for the analogous site percolation
$\epsilon$-expansion).  These values are only indicative as they are
obtained from the leading terms of a series which (like the
${\epsilon}$-expansion series for the other exponents in ISGs) has
never been summed. However, the values are qualitatively consistent
with the observed $\theta(d)$ and $\omega(d) = \theta(d)/\nu(d)$
values for $d= 3$, $4$ and $5$. FSS estimates in $3$d are $\omega =
1.12(10)$ and $\nu = 2.56(4)$ \cite{baity:13} so $\theta(3) =
\omega\nu \approx 3$.  HTSE estimates for various ISG models in $4$d
are $\theta(4) \approx 1.5$ and in $5$d, $\theta(5) \approx 1.0$
\cite{daboul:04,klein:91}. Simulation estimates in $4$d and $5$d are
consistent with these values \cite{banos:12,lundow:14,lundow}.
Potentially there can also be an analytic correction with exponent
\cite{ballesteros:99} $\omega_{a} \approx \gamma/\nu > 2$ in $4$d. If
an analytic contribution is indeed present in $4$d in addition to the
leading conformal correction, the effective exponent of the combined
correction will be slightly increased.

The relative strengths of the leading conformal correction and the
analytic correction will depend on the observable so the effective
exponent for the combination can be expected to vary somewhat from
observable to observable within a narrow range of values.

\begin{figure}
  \includegraphics[width=3.5in]{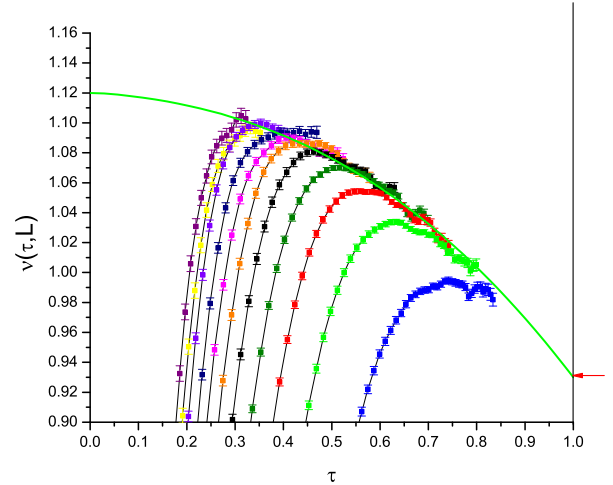}
  \caption{(Color online) The $4$d bimodal ISG. The effective
    correlation length exponent $\nu(\tau,L)$ with
    $\beta_{c}=0.505$. Sizes $L= 14$ to $4$ from left to right. The
    green continuous curve is an overall fit. The red arrow is the
    exact bimodal high temperature limit $\nu(\tau=1) =
    (d-1/3)\beta_{c}^2$.}\protect\label{fig:6}
\end{figure}

\begin{figure}
  \includegraphics[width=3.5in]{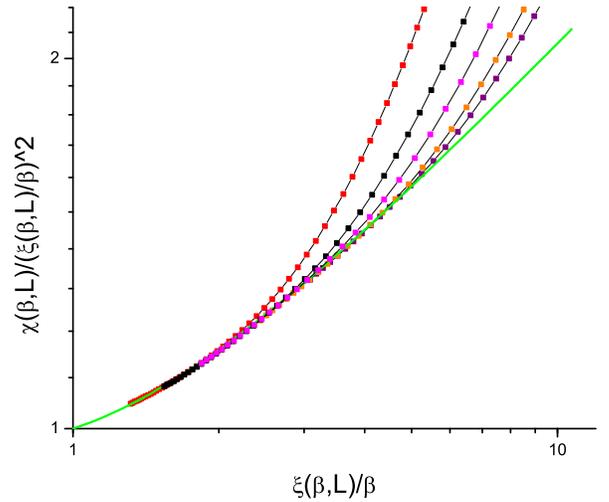}
  \caption{(Color online) The $4$d bimodal ISG. The ratio
    $\chi(\beta,L)/(\xi(\beta,L)/\beta)^2$ as function of
    $\xi(\beta,L)/\beta$ (log-log scaling). Sizes $L=6, 8, 10, 12, 14$
    from left to right. The green continuous curve is an overall
    fit.}\protect\label{fig:7}
\end{figure}

The leading irrelevant scaling field correction is by definition the
correction of this type which has the smallest exponent $\theta$. From
the overall agreement between the $\epsilon$-expansion and observed
values, one can be fully confident that the leading irrelevant
operator correction in $4$d is $\omega(4) \approx 1.5$, and hence no
hypothetical correction term with a much smaller exponent (which if it
existed would modify the behavior at very large $L$) can
exist. Extrapolations of observed FSS data to infinite $L$ using a
correction exponent $\omega \approx 1.3$ are valid, which justifies
the natural straight line extrapolations at $\beta_{c}$ in
Figs.~\ref{fig:1} and \ref{fig:2}.  Similarly valid fits to the ThL
data in Figs.~\ref{fig:6} and \ref{fig:7} are made with a leading
correction term having an exponent $\theta \approx 1.5$. No
hypothetical small exponent correction term producing "reentrant"
behavior exists.

For each dimensionless observable $U(\beta,L)$, extrapolation to
infinite $L$ at $\beta_{c}$ gives an estimate for a critical parameter
$U(\beta_c, \infty)$ characteristic of the universality class. For the
bimodal ISG in dimension four these data show $g(\beta_c,\infty)=
0.525(5)$ and $(\xi/L)(\beta_c,\infty)=0.485(5)$.

With the FSS rule Eq.~\eqref{FSS} for a dimensionless observable
$U(\beta,L)$ the critical exponent $\nu$ can be estimated in principle
from $[dU(\beta,L)/d\beta)]_{\beta_c}= B L^{1/\nu}$. To obtain these
$\nu$ estimates, the fits are multi-parameter as they involve
simultaneous estimates for $\beta_{c}$, $\omega$, and the parameters
$A$, $B$ and $C$ as well as $\nu$. Nevertheless in practice reasonably
precise values for $\nu$ can be obtained. From analyses of data sets
such as those shown in Figs.~\ref{fig:1} and \ref{fig:2} on a number of
different dimensionless observables, a global estimate for the bimodal
model is $\nu = 1.13(1)$ \cite{lundow:14}.

\section{Thermodynamic derivative peak analysis}
\begin{figure}
  \includegraphics[width=3.5in]{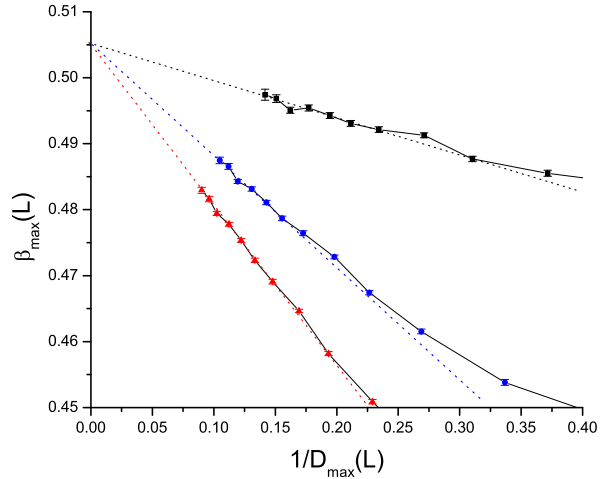}
  \caption{(Color online) The $4$d bimodal ISG. Thermodynamic
    derivative peak $D_{\max}=[\partial U(\beta,L)/\partial
      \beta]_{\max}$ data for dimensionless observables $U=W_{q}, h, g$
    (black squares, blue circles, red triangles). The peak location
    $\beta_{\max}(L)$ against the inverse peak height $1/D_{\max}(L)$.
  }\protect\label{fig:3}
\end{figure}

\begin{figure}
  \includegraphics[width=3.5in]{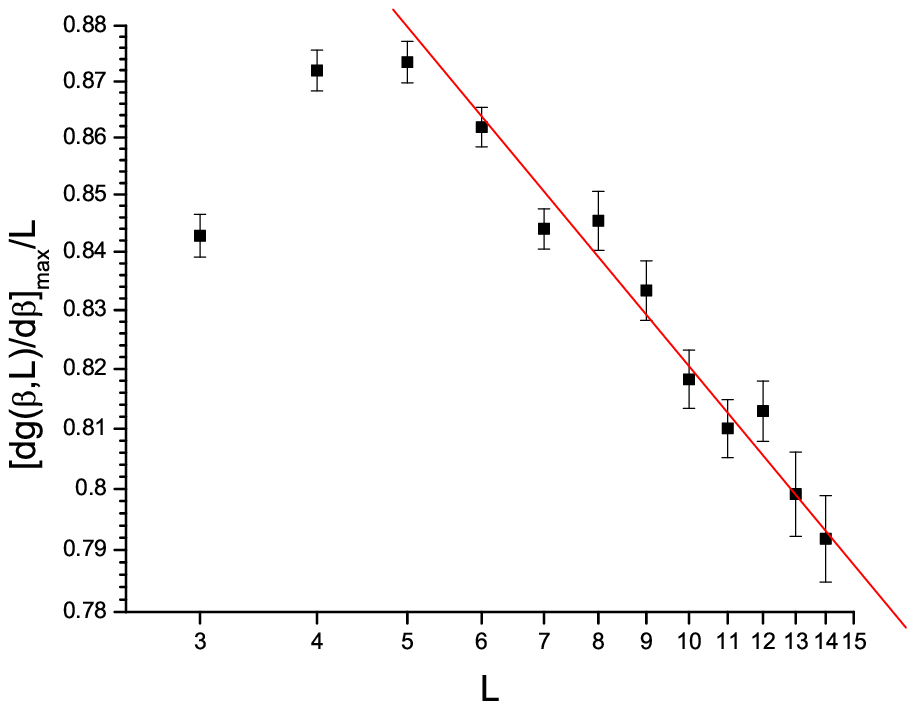}
  \caption{(Color online) The $4$d bimodal ISG. The Binder cumulant
    derivative peak height maximum $[\partial g(\beta,L)/\partial
      \beta]_{\max}$ normalized by $L$, against
    $L$ (log-log scaling).}\protect\label{fig:4}
\end{figure}

\begin{figure}
  \includegraphics[width=3.5in]{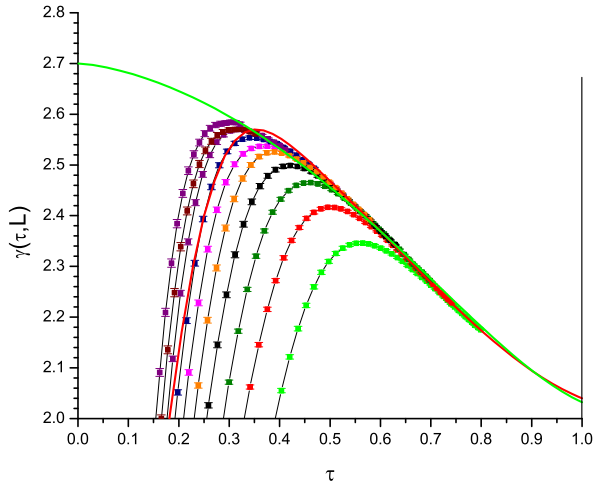}
  \caption{(Color online) The $4$d bimodal ISG. The effective
    susceptibility exponent $\gamma(\tau,L)$ with
    $\beta_{c}=0.505$. Sizes $L= 14$ to $5$ from left to right. The
    red continuous curve is calculated from the tabulated HTSE terms
    in Ref.~\cite{daboul:04}. The green continuous curve is an overall
    fit.  }\protect\label{fig:5}
\end{figure}

Near criticality in a ferromagnet for dimensionless observables
$U(\beta,L)$ the heights of the peaks $D_{\max}(U,L)$ of the
thermodynamic derivatives $D(U,L) =\partial U(\beta,L)/\partial \beta$
scale for large $L$ as \cite{ferrenberg:91,weigel:09}
\begin{equation}
  \lbrack\partial U(\beta,L)/\partial \beta\rbrack_{\max} \propto L^{1/\nu}
\left(1+ aL^{-\omega/\nu}+\cdots \right)
\label{dUdbmax}
\end{equation}
The temperature location of the derivative peak $\beta_{\max}(U,L)$
similarly scales as
\begin{equation}
\beta_{c}-\beta_{\max}(U,L) \propto L^{-1/\nu}(1+bL^{-\omega/\nu} +\cdots)
\label{betamax}
\end{equation}
As both $\beta_{c}-\beta_{\max}(U,L)$ and $1/D_{\max}(U,L)$ vary as
$L^{-1/\nu}$ at large $L$, a plot of the peak locations against the
inverse peak heights tends linearly to $\beta_{c}$ at large $L$.
These estimates of $\beta_{c}$ are independent of the FSS estimates.
The observables used for $U(\beta,L)$ \cite{ferrenberg:91} can be for
instance the Binder cumulant $g(\beta,L)$ or the logarithm of the
finite size susceptibility $\ln(\chi(\beta,L))$.  

For an ISG just the same thermodynamic differential peak methodology
can be used as in the ferromagnet. As far as we are aware this
analysis has not been used previously in the ISG context.  In Fig.~\ref{fig:3}
we show data for peak height locations $\beta_{\max}(U,L)$ against the
inverse peak height $1/D_{\max}(U,L)$ for the derivatives $\partial
U(\beta,L)/\partial \beta$ with dimensionless observables
$U(\beta,L)$, the Binder cumulant $g(\beta,L)$, together with
$W_{q}(\beta,L)$ and $h(\beta,L)$ defined by
\begin{equation}
  W_{q}(\beta,L) = \frac{1}{\pi-2} \left(\pi\frac{\lbrack\langle
    |q|\rangle\rbrack^2}{\lbrack\langle q^2\rangle\rbrack} - 2\right)
  \label{Wqdef}
\end{equation}
and
\begin{equation}
  h(\beta,L) = \frac{1}{\sqrt{\pi}-\sqrt{8}}
  \left(
  \sqrt{\pi}\frac{
    \lbrack\langle|q^{3}|\rangle\rbrack
  }{
    \lbrack\langle q^{2}\rangle\rbrack^{3/2}
  }-\sqrt{8}
  \right)
  \label{hdef}
\end{equation}
where the coefficients have been chosen so that the parameters go from
0 at high temperature to 1 at low temperature.  In Fig.~\ref{fig:3}
the linear extrapolations of the three sets of data points lead
consistently to $\beta_{c}= 0.505(1)$, in full agreement with the FSS
data.

From the thermodynamic derivatives, "The critical exponent $\nu$ can
be estimated without any consideration of the critical coupling
$\beta_{c}$" \cite{ferrenberg:91}.  As the exponent $\nu$ is close to
$1$ in $4$d, we choose to plot data for the normalised Binder cumulant
derivative $\log(D_{\max}(L)/L)$ against $\log(L)$, Fig.~\ref{fig:4}.
Corrections to scaling are visible for small $L$, but for $L > 4$ the
slope is $-0.0890(5)$, which corresponds to $\nu = 1.13(1)$, in full
agreement with the FSS estimate above.

\section{Thermodynamic limit derivatives}
Analyses using the standard RGT scaling variable $t=1- T/T_{c}=
1-\beta_{c}/\beta$, either in ferromagnets or in spin glasses, are
restricted to the critical region and the FSS regime $L \ll
\xi(\beta,\infty)$, because $t$ tends to diverge at high temperatures
so scaling corrections automatically proliferate outside the critical
regime. This is unnecessary.  A natural scaling variable, which has
been used for the scaling of the susceptibility in ferromagnets for
more than 50 years, is $\tau = 1-\beta/\beta_{c}$
\cite{fisher:67,wegner:72,gartenhaus:88,butera:02}. The Wegner
thermodynamic limit (ThL) susceptibility scaling expression
\cite{wegner:72}, which is expressed in terms of $\tau$ and is valid
from criticality to infinite temperature,is
\begin{equation}
  \chi(\tau,\infty) = C_{\chi}\tau^{-\gamma}\left(1 + a_{\chi}\tau^{\theta} + b_{\chi}\tau  + \cdots\right)
  \label{wegner}
\end{equation}
At criticality $\tau$ becomes identical to $t$ and tends to $1$ at
infinite temperature, so the corrections in the Wegner expression are
well behaved for the entire paramagnetic temperature range. The first
correction term in the equation is the leading confluent correction
and the second an analytic correction. There is a closure condition as
$\chi(\tau=1) \equiv 1$.  In ISGs as the interaction energy is
$\langle J^2 \rangle$, not $\langle J \rangle$, and the appropriate
scaling variable is $\tau = 1-(\beta/\beta_c)^2$ or $\tau_{w} = 1
-(\tanh(\beta)/\tanh(\beta_{c}))^2$. This is the scaling variable
which was used from the beginning in ISG HTSE work
\cite{fisch:77,singh:86,klein:91,daboul:04,campbell:06} but curiously
not in most analyses of numerical simulations. The Wegner expression
for the susceptibility \cite{wegner:72} carries over, with the
appropriate definition for $\tau$.

It has been pointed out that in Ising ferromagnets the ThL expression
for the second moment correlation length $\xi(\tau)$ analogous to the
Wegner expression for the susceptibility is
\cite{campbell:06,campbell:08,butera:02,lundow}
\begin{equation}
  \xi(\tau) = C_{\xi}\beta^{1/2}\tau^{-\nu}\left(1+ a_{\xi}\tau^\theta + \cdots\right)
\label{xiferro}
\end{equation}
The factor $\beta^{1/2}$ arises because the generic infinite
temperature limit behavior is $\xi(\tau)/\beta^{1/2} \to 1$.

In ISGs the factor $\beta^{1/2}$ becomes $\beta$, again because
$\langle J^2 \rangle$ replaces $\langle J \rangle$ \cite{campbell:06},
so
\begin{equation}
  \xi(\tau) = C_{\xi}\beta\tau^{-\nu}\left(1+ a_{\xi}\tau^\theta + \cdots\right)
\label{xiISG}
\end{equation}
with the ISG $\tau$.  Following a well-established protocol
\cite{kouvel:64,butera:02} one can define a temperature dependent
effective exponent for the susceptibility
\begin{equation}
  \gamma(\tau)= -\partial\ln(\chi(\tau))/\partial\ln(\tau)
\label{gamtau}
\end{equation}
and for the second moment correlation length
\begin{equation}
  \nu(\tau)= -\partial\ln(\xi(\tau)/\beta)/\partial\ln(\tau)
\label{nutau}
\end{equation}
which tend to the critical $\gamma$ and $\nu$ respectively at
criticality, and to $2d\beta_{c}^2$ and $(d-K/3)\beta_{c}^2$
respectively in the high temperature $\tau = 1$ limit, where $K$ is
the kurtosis of the interaction distribution. ($K =1$ in the bimodal
case and $K=3$ for the Gaussian).

For data in the ThL regime, the Wegner expression with two correction
terms Eq.~\eqref{wegner} translates exactly into
\begin{equation}
  \gamma(\tau) = \gamma - \frac{a_{\chi}\theta\tau^{\theta}+b_{\chi}y\tau^{y}}{1+a_{\chi}\tau^{\theta}+b_{\chi}\tau^{y}}
\label{gamtauWeg}
\end{equation}
where the second term is an effective subleading correction.  There is
an equivalent relation for $\nu(\tau)$.  Fixing $\beta_{c} = 0.505$,
the $\gamma(\tau)$ and $\nu(\tau)$ plots for all the different sizes
$L$ are shown in Fig.~\ref{fig:5} and Fig.~\ref{fig:6}. In
Fig.~\ref{fig:5} the effective $\gamma(\beta)$ evaluated from an
explicit sum of the first 15 terms in the HTSE series of
Ref.~\cite{daboul:04} is also shown.

The fit curves are adjusted to those data which are in the ThL regime,
and the fit is then extrapolated to criticality to obtain estimates
for the critical $\gamma$ and $\nu$. Just as for $\omega$ in the FSS
regime, $\theta$ is the lowest correction exponent and so a further
hypothetical correction term with a much lower exponent which could
modify the form of the fit curve between the ThL data and criticality
can be ruled out. The $\gamma(\tau)$ and $\nu(\tau)$ plots are however
rather sensitive to the exact value of $\beta_{c}$.

The fits shown correspond to an ISG temperature dependent
susceptibility effectively fitted by
\begin{equation}
\chi(\beta) = 0.634\tau^{-2.70}\left(1+0.38\tau^{1.5}+0.22\tau^{3}-0.022\tau^{8}\right)
\label{Jgamfit}
\end{equation}
and a second moment correlation length
\begin{equation}
\xi(\beta) = 0.914\beta\tau^{-1.12}\left(1+0.055\tau^{1.5}+0.042\tau^{3}\right)
\label{Jnufit}
\end{equation}
so with estimated critical exponents $\gamma = 2.70$, $\nu = 1.12$,
$\theta = 1.5$, with a leading correction term and further higher
order terms.

The ratio 
\begin{equation}
  \frac{\gamma(\tau,L)}{\nu(\tau,L)} = \frac{\partial \ln\chi(\beta^2,L)/
  \partial \ln\tau}{\partial \ln(\xi(\beta^2,L)/\beta)/ \partial
  \ln\tau} =2-\eta(\beta,L)
\end{equation}
does not involve $\beta_c$, so $\eta(\beta,L)$ can be measured without
any knowledge of $\beta_{c}$. Then the $\eta(\beta,L)$ as functions of
$\beta/\xi(\beta,L)$ can be extrapolated to $\beta/\xi(\beta,L)=0$ for
infinite $L$ to estimate the critical value of $\eta$ without
involving $\beta_{c}$.
Alternatively we plot, see Fig.~\ref{fig:7},
$y=\chi(\beta,L)/(\xi(\beta,L)/\beta)^2$ against
$x=\xi(\beta,L)/\beta$. At infinite temperature $y=x=1$, and towards
criticality for the ThL regime data $y \propto \tau^{(2-\eta)\nu-2\nu}
= \tau^{-\nu\eta}$ and $x \propto \tau^{-\nu}$. Hence the slope of
$\log(y)$ against $\log(x)$ tends to $-\eta$ as criticality is
approached. The fit and limiting slope can be estimated without
specifying an explicit value of $\beta_{c}$.  A satisfactory fit curve
with a single effective correction term is
$y=Cx^{-\eta}(1+(C^{-1}-1)x^{-\theta})$ with $\eta$, $\theta$ and $C$
adjustable parameters. The values in the fit shown are $\eta =
-0.415(10)$, $\theta=1.3$, and $C=0.78$.

The critical exponents estimated without any knowledge of $\beta_{c}$
are therefore $\nu = 1.13(1)$ from the thermodynamic derivative result
Fig.~\ref{fig:4} and $\eta = -0.415(10)$, and so $\gamma = \nu(2-\eta)
= 2.73(3)$ combining these values. The values are in excellent
agreement with the estimates from the extrapolated ThL derivatives of
Fig.~\ref{fig:5} and Fig.~\ref{fig:6} which validates the $\beta_{c}$
estimate and the overall methodology.

\section{High temperature series expansion}
\begin{figure}
  \includegraphics[width=3.5in]{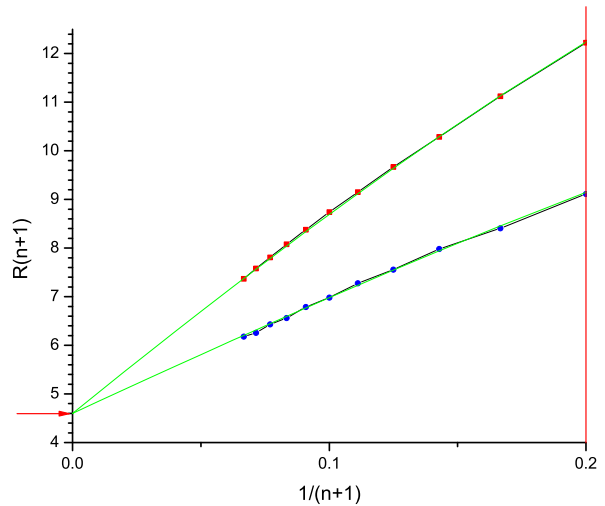}
  \caption{(Color online) The $4$d bimodal ISG. The HTSE ratio series
    $R(n+1)$ for the higher order susceptibilities $\Gamma_{4}$ (red
    squares) and $\Gamma_{3}$ (blue circles). Green curves :
    fits. Arrow : critical point.}\protect\label{fig:8}
\end{figure}
In the HTSE approach, series of exact terms are evaluated, which are
summed to obtain the temperature dependence of the susceptibility. In
ISGs the series for the spin glass susceptibility is
\begin{equation}
\chi(w) = 1 + a(1)w+ a(2)w^2 + a(3)w^3 + \cdots
\end{equation}
where $w = \tanh^2(\beta)$ \cite{singh:86,klein:91}, or the same
equation with $w$ replaced by $\beta^2$ \cite{daboul:04}. The terms
$a(n)$ are exact.  The number of terms which can be calculated is
limited by practical considerations, and up to now $n=15$ has been an
upper limit in ISGs. The series have been analysed by Dlog-Pad\'e, M1
and M2 techniques \cite{singh:86,klein:91,daboul:04} which are not
particularly transparent for a non-specialist. An alternative
classical approach is the ratio method
\cite{domb:56,fisher:67,butera:02}. For a model where the temperature
dependence of the thermodynamic limit (infinite size) susceptibility
is \cite{wegner:72}
\begin{equation}
\chi(w) = C_{\chi}\tau^{-\gamma}\left(1 + a_{\chi}\tau^{\theta}+\cdots\right)
\end{equation}
with $\tau = 1 -w/w_{c}$, the corresponding HTSE terms are \cite{butera:02}
\begin{equation}
  a(n)= \frac{Cn^{\gamma-1}}{\Gamma(\gamma) w_c^n}\left(1+\frac{\Gamma(\gamma)}{\Gamma(\gamma-\theta)}\frac{a_{\chi}}{n^{\theta}}+ \cdots\right)
  \label{anbutera}
\end{equation}
The ratios between successive terms are
\begin{multline}
  R(n+1)=\frac{a(n+1)}{a(n)}=  \\
  = \frac{1}{w_{c}}\left(1+\frac{\gamma-1}{n+1}\right)
  \left(1 -\frac{\Gamma(\gamma)}{\Gamma(\gamma-\theta)}
  \frac{\theta a_{\chi}}{(n+1)^{\theta+1}}\right)
\label{anratio}
\end{multline}
In favorable cases the ratio series $R(n+1)$ plotted as a function of
$1/(n+1)$ can be extrapolated to infinite $n$ to estimate the critical
temperature from the intercept, $1/w_c$, the critical exponent
$\gamma$ from the initial slope $\partial R(n+1)/\partial(1/(n+1)) =
(\gamma-1)/w_{c}$, and the conformal correction exponent $\theta$ from
the leading non-linear term.  Unfortunately there can be parasitic
oscillating terms in the ratios, arising from anti-ferromagnetic poles
\cite{butera:02}, particularly in simple hypercubic lattices.  It
turns out that in the ISGs the parasitic terms in the susceptibility
series become rather strong when the dimension drops. Explicit lists
of terms for $\Gamma_{2}$ in dimension $d=4$ and below were tabulated
in Ref.~\cite{singh:86}.

For the simple hypercubic bimodal ISGs in general dimension Klein {\it
  et al.} \cite{klein:91} tabulated the terms to order 15 not only for
the ISG susceptibility $\chi(w)$ series (called $\Gamma_{2}$ in their
nomenclature) but also for the higher order susceptibilities
$\Gamma_{3}$ and $\Gamma_{4}$ which they define. These
susceptibilities have critical exponents $\gamma_{3} = (3\gamma
+d\nu)/2$ and $\gamma_{4} = 2\gamma + d\nu$ respectively
\cite{klein:91,butera:02}.  Unfortunately the parasitic terms are so
strong in the $4$d $\Gamma_{2}$ susceptibility series that the ratios
cannot be readily exploited so as to make accurate estimates of the
critical parameters. This explains why the error bars quoted in
Ref.~\cite{daboul:04} for $\beta_{c}^2$ and $\gamma$ in $4$d are
rather large. However, we have summed the tabulations of
\cite{klein:91} for $\Gamma_{3}$ and $\Gamma_{4}$ to obtain the
$a_{3}(n)$ and $a_{4}(n)$ for these series, and have found that for
these higher order susceptibilities the ratios behave very regularly
in dimension $4$ (and above), as can be seen in Fig.~\ref{fig:8}.
Expression \ref{anratio} can be applied to these ratios and reliable
extrapolations can be made through which one can estimate $w_c,
\gamma(3), \gamma(4)$ and $\theta$. Optimal fits excluding the small
$n$ values are
\begin{equation}
  R(n+1) = 4.60\left(1 +\frac{5.3}{(n+1)}- \frac{4.0}{(n+1)^{2.35}}\right)
\label{Gam3fit}
\end{equation}
for $\Gamma_{3}$, and
\begin{equation}
  R(n+1) = 4.60\left(1 +\frac{9.15}{(n+1)}- \frac{9.0}{(n+1)^{2.4}}\right)
\label{Gam4fit}
\end{equation}
for $\Gamma_{4}$.  The joint intercept $1/w_{c} = 1/\tanh^2(\beta_{c})
= 4.60(2)$ corresponds to $\beta_{c}=0.505(2)$, and the initial slopes
correspond to $\gamma_{3} = (3\gamma +d\nu)/2 = 6.30(10)$ and
$\gamma_{4} = 2\gamma + d\nu = 10.15(10)$.  The leading correction
terms have an exponent $\theta \approx 1.4$, and prefactors with
negative sign. By inspection any hypothesis of another term (which
would then be the true conformal correction to scaling) having a much
lower exponent and with a prefactor of the opposite sign can once
again be ruled out.  There are no statistical errors as all points are
exact; systematic errors arising from the extrapolation are small.

The exponents estimated above from the simulations and so quite
independently from the HTSE data correspond to $\gamma_{3}=(3\gamma
+4\nu)/2 = 6.3(1)$ and $\gamma_{4}= 2\gamma + 4\nu = 9.92(15)$, and so
are fully consistent with the HTSE values.

\section{Gaussian interaction model}
We will present equivalent data for the hypercubic ISG model with
Gaussian interactions in dimension four. All the discussions above
apply equally well to the Gaussian data. Corrections for the Gaussian
model are in general much weaker than for the bimodal
model. Unfortunately no HTSE higher order susceptibility data are
available for the Gaussian model.

\begin{figure}
  \includegraphics[width=3.5in]{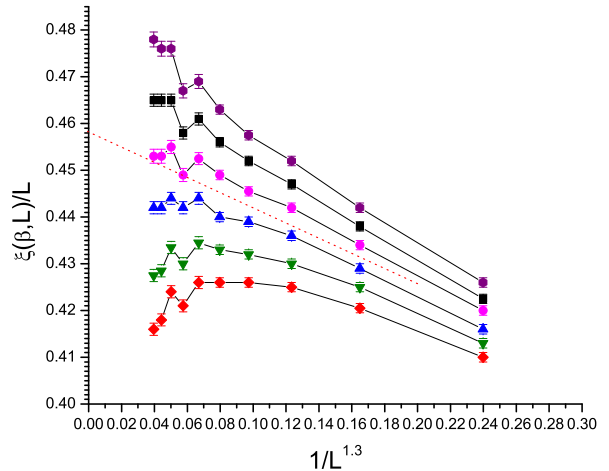}
  \caption{(Color online) The $4$d Gaussian ISG. FSS for the
    correlation length ratio $\xi(\beta,L)/L$ at fixed inverse
    temperatures $\beta=0.5625$, $0.5600$, $0.5575$, $0.5550$,
    $0.5525$, $0.5500$  from top to bottom, against
    $1/L^{1.3}$. Dashed line : critical
    behavior. }\protect\label{fig:9}
\end{figure}

\begin{figure}
  \includegraphics[width=3.5in]{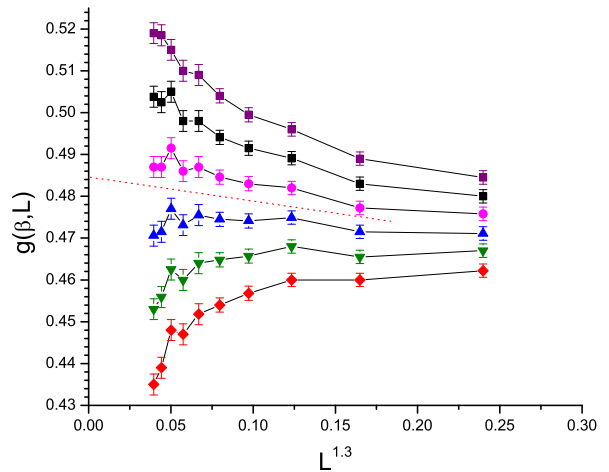}
  \caption{(Color online) The $4$d Gaussian ISG. FSS for the Binder
    cumulant $g(\beta,L)$ at fixed inverse temperatures $\beta=0.5625$,
    $0.5600$, $0.5575$, $0.5550$, $0.5525$, $0.5500$ from top to
    bottom, against $1/L^{1.3}$. Dashed line : critical
    behavior. }\protect\label{fig:10}
\end{figure}

\begin{figure}
  \includegraphics[width=3.5in]{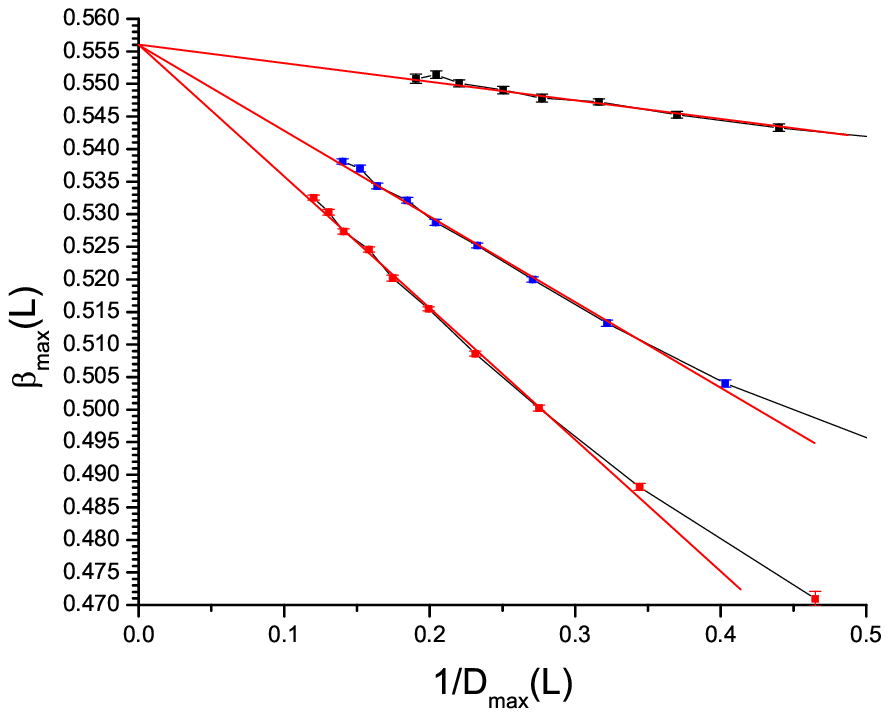}
  \caption{(Color online) The $4$d Gaussian ISG. Thermodynamic
    derivative peak $D_{\max}=[\partial U(\beta,L)/\partial
      \beta]_{\max}$ data for dimensionless observables $U=W_{q}, h, g$
    (black squares, blue circles, red triangles). The peak location
    $\beta_{\max}(L)$ against the inverse peak height $1/D_{\max}(L)$.
  }\protect\label{fig:11}
\end{figure}

\begin{figure}
  \includegraphics[width=3.5in]{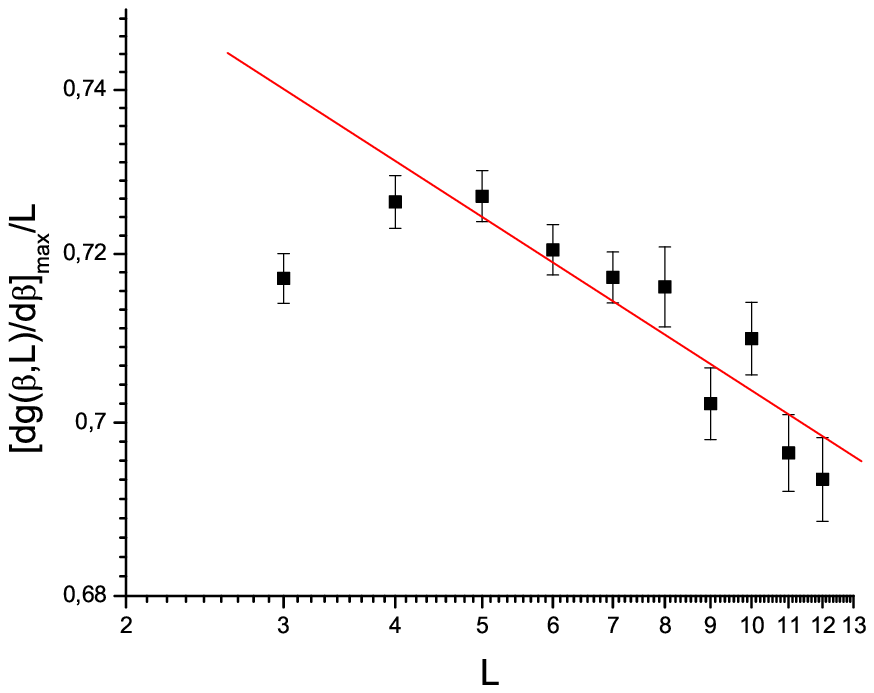}
  \caption{(Color online) The $4$d Gaussian ISG. The Binder cumulant
    derivative peak height maximum $[\partial g(\beta,L)/\partial
      \beta]_{\max}$ normalized by $L$, against $L$ (log-log
    scaling). }\protect\label{fig:12}
\end{figure}

\begin{figure}
  \includegraphics[width=3.5in]{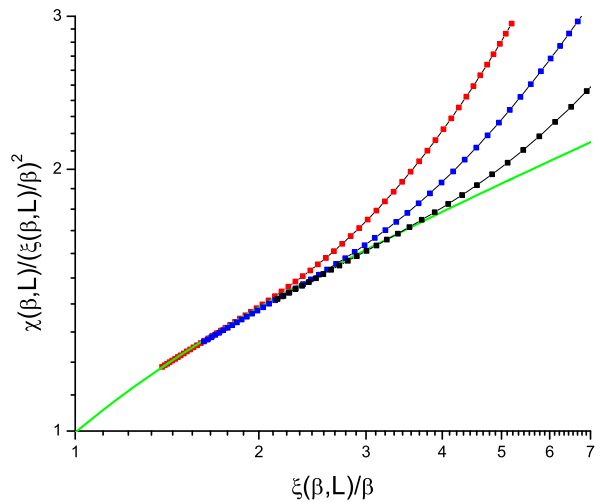}
  \caption{(Color online) The $4$d Gaussian ISG. The ratio
    $\chi(\beta,L)/(\xi(\beta,L)/\beta)^2$ as function of
    $\xi(\beta,L)/\beta$ (log-log scaling). Sizes $L= 6, 8, 12$ from
    left to right.  }\protect\label{fig:13}
\end{figure}

\begin{figure}
  \includegraphics[width=3.5in]{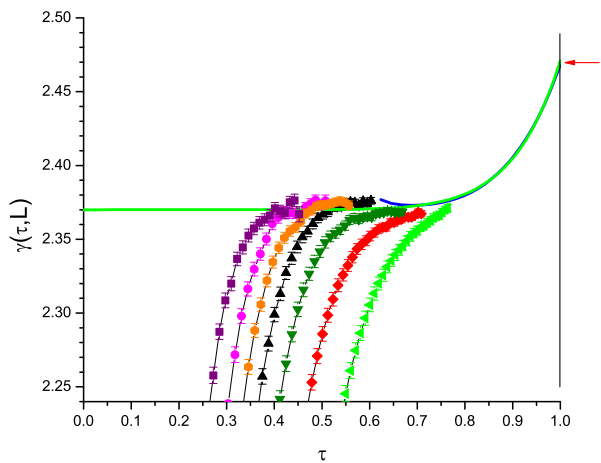}
  \caption{(Color online) The $4$d Gaussian ISG. The effective
    exponent $\gamma(\tau)$ with $\beta_{c}=0.5555$.  Sizes $L= 12,
    10, 9, 8, 7, 6, 5$ from left to right. The blue continuous curve
    is calculated from the tabulated HTSE terms in
    Ref.~\cite{daboul:04}.  The red arrow is the exact high
    temperature limit $\gamma(\tau=1) = 2d\beta_{c}^2$.
  }\protect\label{fig:14}
\end{figure}

\begin{figure}
  \includegraphics[width=3.5in]{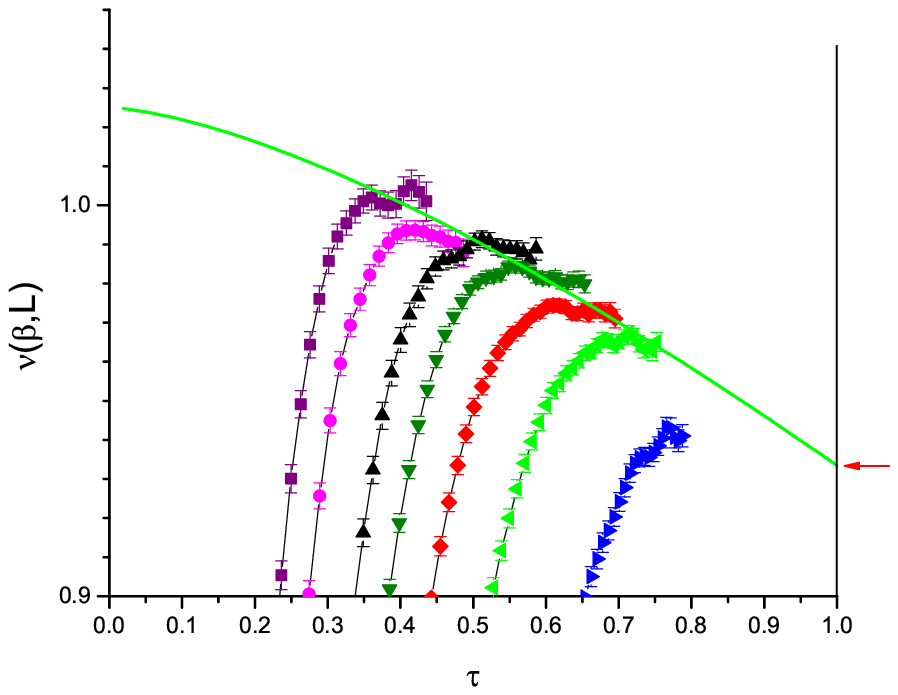}
  \caption{(Color online) The $4$d Gaussian ISG. The effective
    exponent $\nu(\tau)$ with $\beta_{c}=0.5555$. Sizes $L= 12, 10, 8,
    7, 6, 5, 4$ from left to right. The red arrow is the exact
    Gaussian high temperature limit $\nu(\tau=1) = (d-1)\beta_{c}^2$.
  }\protect\label{fig:15}
\end{figure}

The $4$d Gaussian critical inverse temperature from both the FSS data,
Figs.~\ref{fig:9} and \ref{fig:10}, and the thermodynamic derivative
peak locations Fig.~\ref{fig:11}, is estimated to be $\beta_{c} =
0.5555(10)$ or $T_{c}=1.800(3)$. This value is fully consistent with
previous FSS estimates from Binder parameter and correlation length
ratio measurements $0.555(3)$ \cite{parisi:96,ney:98} and
0.554(3)\cite{jorg:08}.

The Gaussian critical parameters for the normalised second moment
correlation length $\xi(\beta,L)/L$ and the Binder cumulant
$g(\beta,L)$ from Figs.~\ref{fig:9} and \ref{fig:10} are $(\xi/L)_{c}
= 0.458(5)$ and $g_{c}= 0.485(5)$ respectively. The value for the
Binder cumulant is close to the estimate $g_{c}= 0.470(5)$ given in
Ref.~\cite{jorg:08}.

The exponent $\nu$ estimated directly form the Binder cumulant
thermodynamic derivative peak heights, Fig.~\ref{fig:12}, is
$1.030(5)$. This value is fully consistent with the estimate
$\nu=1.02(2)$ given in Ref.~\cite{jorg:08}.

The plot of $\log(y)$ against $\log(x)$ with $y =
\chi(\beta,L)/(\xi(\beta,L)/\beta)^2$ and $x = \xi(\beta,L)/\beta$,
Fig.~\ref{fig:13}, is fitted with a single effective correction :
\begin{equation}
  y(x) = 1.2x^{0.305}\left(1-0.17x^{-1.4}\right)
\end{equation}
corresponding to a critical exponent $\eta = -0.305(10)$. This value
is again fully consistent with but more accurate than the estimate
$\eta =-0.275(25)$ given by \cite{jorg:08}.

The fit to the ThL effective $\gamma(\tau)$ with a fixed
$\beta_{c}=0.5555$, Fig.~\ref{fig:14}, corresponds to
\begin{equation}
  \chi = 1.01\tau^{-2.37}\left(1-0.010\tau^{10}\right)
\end{equation}
so with a critical exponent $\gamma = 2.37(2)$ and only a very weak
high order correction term. The prefactor for the usual leading
correction term appears to be accidentally very close to zero. A
similar conclusion was reached from the FSS analysis of the $4$d
Gaussian model in \cite{jorg:08}.  The fit to the ThL effective
$\nu(\tau)$ with fixed $\beta_{c}=0.5555$, Fig.~\ref{fig:15},
corresponds to
\begin{equation}
  \xi(\beta)= 0.934\tau^{-1.025}\beta\left(1+0.070\tau^{1.5}\right)
\end{equation}
so a critical exponent $\nu = 1.025(3)$ and a correction term with a
similar exponent to the exponent of the bimodal model leading
correction.

The critical constants and exponents for a $4$d diluted bimodal model
Ref.~\cite{jorg:08} were estimated to be $g(\beta_c,\infty)=
0.472(2)$, $(\xi/L)(\beta_c,\infty) \approx 0.440 $, $\gamma = 2.33(6)$,
and $\nu = 1.025(15)$. Each of the diluted bimodal critical values is
similar to that for the $4$d Gaussian model, but the diluted bimodal
model and the standard bimodal model studied above have quite
different critical properties.


\section{Conclusion}
We compare in Table~\ref{Table:I} the bimodal and Gaussian interaction
distribution model estimates for each of the critical parameters and
critical exponents studied above.  For all the parameters and
exponents the estimates for the two models are quite different, with
in each case differences of the order of 10\%.
\begin{table}[htbp]
  \caption{\label{Table:I} Values of the $4$d bimodal and Gaussian
    critical parameters and exponents. }
  \begin{ruledtabular}
    \begin{tabular}{ccc}\\
      Parameter& Bimodal & Gaussian \\
      \hline
      $\beta_c$ & $0.505(1)$ & $0.5555(10)$ \\
      $g(\beta_c,\infty)$&$0.525(5)$&$0.485(5)$ \\
      $(\xi/L)(\beta_c,\infty)$&$0.485(5)$&$0.458(5)$ \\
      $\gamma$&$2.70(3)$&$2.37(2)$\\
      $\nu$&$1.113(1)$&$1.030(5)$\\
      $\eta$&$0.42(2)$&$0.305(10)$
    \end{tabular}
  \end{ruledtabular}
\end{table}
We conclude that in $4$d ISGs the critical exponents, like the values
of the critical ordering parameters, depend on the form of the
interaction distribution. ISGs in dimension four with different
interaction distributions are clearly not in the same universality
class.

\section{Acknowledgements}
The simulations were performed on resources provided by the Swedish
National Infrastructure for Computing (SNIC) at High Performance
Computing Center North (HPC2N) and at Chalmers Centre for
Computational Science and Engineering (C3SE).

\end{document}